\documentclass[prl,twocolumn,superscriptaddress,showpacs,aps]{revtex4}
\usepackage{graphicx}
\usepackage{amssymb}
\usepackage{amsmath}
\usepackage{amsfonts}
\usepackage{dcolumn}

\input psfig.sty

\begin{document}
\preprint{UVA-INPP-02-01}
\title{Precise Measurement of the Pion Axial Form Factor 
in the {\boldmath $\pi^+\to e^+\nu\gamma$} Decay} 

\newcommand*{\uva}{Department of Physics, University of Virginia,
                   Charlottesville, VA 22904-4714, USA} 
\newcommand*{\psii}{Paul Scherrer Institut, Villigen PSI, CH-5232,
                    Switzerland}
\newcommand*{\dubna}{Joint Institute for Nuclear Research, RU-141980
                     Dubna, Russia}
\newcommand*{\swierk}{Institute for Nuclear Studies, PL-05-400 Swierk,
                      Poland}
\newcommand*{\tbilisi}{Institute for High Energy Physics, Tbilisi
                       State University, GUS-380086 Tbilisi, Georgia}
\newcommand*{\asu}{Department of Physics and Astronomy, Arizona
                   State University, Tempe, AZ 85287, USA}
\newcommand*{\irb}{Rudjer Bo\v{s}kovi\'c Institute, HR-10000 Zagreb,
                   Croatia}

\affiliation{\uva}
\affiliation{\psii}
\affiliation{\dubna}
\affiliation{\swierk}
\affiliation{\tbilisi}
\affiliation{\asu}
\affiliation{\irb}

\author{E.~Frle\v{z}}\email[Corresponding author: ]{frlez@virginia.edu}\affiliation{\uva}
\author{D.~Po\v{c}ani\'c}\affiliation{\uva}
\author{V.~A.~Baranov}\affiliation{\dubna}
\author{W.~Bertl}\affiliation{\psii}
\author{M.~Bychkov}\affiliation{\uva}
\author{N.~V.~Khomutov}\affiliation{\dubna}
\author{A.~S.~Korenchenko}\affiliation{\dubna} 
\author{S.~M.~Korenchenko}\affiliation{\dubna}
\author{T.~Kozlowski}\affiliation{\swierk}
\author{N.~P.~Kravchuk}\affiliation{\dubna}
\author{N.~A.~Kuchinsky}\affiliation{\dubna} 
\author{W.~Li}\affiliation{\uva}
\author{R.~C.~Minehart}\affiliation{\uva}
\author{D.~Mzhavia}\affiliation{\dubna} 
\author{B.~G.~Ritchie}\affiliation{\asu}
\author{S.~Ritt}\affiliation{\psii}
\author{A.~M.~Rozhdestvensky}\affiliation{\dubna} 
\author{V.~V.~Sidorkin}\affiliation{\dubna}
\author{L.~C.~Smith}\affiliation{\uva}
\author{I.~Supek}\affiliation{\irb} 
\author{Z.~Tsamalaidze}\affiliation{\tbilisi}
\author{B.~A.~VanDevender}\affiliation{\uva}
\author{E.~P.~Velicheva}\affiliation{\dubna}
\author{Y.~Wang}\affiliation{\uva}
\author{H.-P.~Wirtz}\altaffiliation[Presently at: ]{Philips 
Semiconductors AG, CH-8045  Z\"urich, Switzerland}\affiliation{\psii}
\author{K.~O.~H.~Ziock}\affiliation{\uva}
\date{9 December 2003}
\begin{abstract}
We have studied radiative pion decays $\pi^+\to e^+\nu\gamma$ in
three broad kinematic regions using the PIBETA detector and a stopped
pion beam.  Based on Dalitz distributions of 42,209 events we have
evaluated absolute $\pi\to e\nu\gamma$ branching ratios in the three
regions.  Minimum $\chi^2$ fits to the integral and differential
$(E_{e^+},E_\gamma)$ distributions result in the axial-to-vector weak
form factor ratio of $\gamma\equiv F_A/F_V=0.443(15)$, or
$F_A=0.0115(4)$ with $F_V=0.0259$.  However, deviations from Standard
Model predictions in the high-$E_\gamma$/low-$E_{e^+}$ kinematic
region indicate the need for further theoretical and experimental
work.
\end{abstract}


\maketitle

In the Standard Model description of radiative pion decay $\pi^+\to
e^+\nu\gamma$, where $\gamma$ is a real or virtual photon ($e^+e^-$
pair), the decay amplitude $M$ depends on the vector $V$ and axial
vector $A$ weak hadronic currents~\cite{Bry82}.  Both currents
contribute to the structure-dependent terms $SD_V$ and $SD_A$
associated with virtual hadronic states, while only the axial-vector
current contributes to the inner bremsstrahlung process $IB$.  Thus,
it is convenient to write the decay amplitude as a sum:
$M=M_{IB}+M_{SD}$.  The $IB$ amplitude is~\cite{Ber58,Kin59}
\begin{align}
   M_{IB}  =& 
    -i \frac{eG_FV_{ud}}{\sqrt{2}} 
         f_\pi m_e\epsilon^{\mu *}\bar{e} 
       \left( \frac{k_\mu}{kq} - \frac{p_\mu}{pq} 
             + \frac{\sigma_{\mu\nu}q^\nu}{2kq} \right) \nonumber \\ 
    & \times \left( 1 - \gamma_5\right) \nu\,,
                                                \label{Eq:2}
\end{align}
where $p$, $k$ and $q$ are the pion, electron and photon four-momenta,
respectively, $e$ and $m_e$ are the electron charge and mass, $G_F$ is
the Fermi coupling constant, $V_{ud}$ is the CKM
quark mixing matrix element, while $f_\pi$ is  the pion decay constant.

The structure-dependent amplitude is parameterized by the vector and
axial vector form factors, $F_V$ and $F_A$: 
\begin{align}
  M_{SD} =& \frac{eG_FV_{ud}}{m_\pi\sqrt{2}}\epsilon^{\nu *} 
             \bar{e}\gamma^\mu(1-\gamma_5)\nu     \nonumber \\
    &  \times \left[ F_V\epsilon_{\mu\nu\sigma\tau} p^\sigma q^\tau
        + i F_A (g_{\mu\nu}pq - p_\nu q_\mu)\right] \,.
                                                      \label{Eq:3}
\end{align}
The conserved vector current (CVC) hypothesis~\cite{Ger55,Fey58}
relates $F_V$ to the $\pi^0$ lifetime 
yielding $F_V=0.0259(5)$~\cite{PDG}, which agrees with the
relativistic quark model and chiral perturbation theory~\cite{Gen03}.
Chiral symmetry calculations~\cite{Hol86,Bij97,Gen03} yield $F_A$ in
the range 0.010--0.014.

The combined $\pi\to e\nu\gamma$ event count of all previously
published experiments is less than 1,200 events, while the overall
uncertainties of the parameter $\gamma\equiv F_A/F_V$ extracted from
data range from 12\,\% to
56\,\%~\cite{Dep63,Ste78,Bay86,Pii86,Dom88,Bol90}.

In this Letter we present a first analysis of the $\pi^+\to
e^+\nu\gamma$ ($\pi e2\gamma$) events recorded with the PIBETA
detector in the course of a new measurement of the $\pi^+\to
\pi^0e^+\nu$ ($\pi\beta$) branching ratio~\cite{Poc88,Poc03} from
1999 to 2001. 

\begin{figure} [b] 
\includegraphics[scale=0.75]{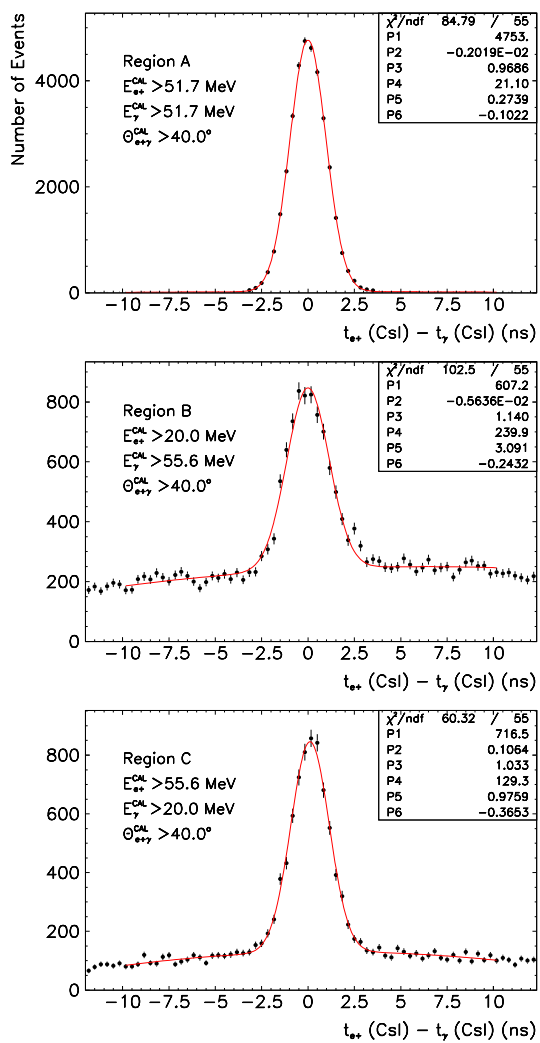}
\caption{The $\pi^+\to e^+\nu\gamma$ class $A$ events (the top
panel) taken with two-arm calorimeter trigger are virtually
background-free (save for the $\pi\beta$ contamination). The
signal-to-background ratio for class $B$ and class $C$ events recorded
with one-arm trigger is 3.8:1 and 7.6:1, respectively.}
\label{rp_sn}
\end{figure}

The measurements were performed in the $\pi E1$ channel at the Paul
Scherrer Institute (PSI), Villigen, Switzerland.  The $\pi^+$ beam
with $p\simeq 113\,$MeV/c, $\Delta p/p\le 1.3\,$\%, and 24 mr
horizontal and vertical divergence, had an average intensity of
$6.8\cdot 10^5\,$$\pi^+$/s, and produced a $\sigma_{x,y}=9\,$mm beam
spot on target.  A total of $2.2\cdot 10^{13}$ $\pi^+$ stops were
recorded during the running period.

The beam particles were first registered in a 3$\,$mm thick beam
counter (BC) placed $\sim$395\,cm upstream of the detector center.
The pions were slowed in a 30\,mm thick active degrader (AD) and
stopped in a 50\,mm long segmented active target detector (AT)
positioned in the center of the PIBETA detector.  The BC, AD and AT
detectors are all made of plastic scintillation material.

The $e^+$ and $\mu^+$ beam contaminations determined by the
time-of-flight method were small, 0.4$\,$\% and 0.2$\,$\%,
respectively.  The background from hadronic interactions was
efficiently suppressed offline by cuts on the timing differences
between the beam detectors and the CsI calorimeter.

The heart of the PIBETA detector is a spherical, 240-element, pure CsI
electromagnetic calorimeter, 12 radiation lengths thick.  The
calorimeter achieves an rms energy resolution of $\sigma_E/E=4.0\,$\%
for $\pi^0$ mass reconstruction at rest, sub-ns timing resolution, and
an angular resolution of $2.0^\circ$ for minimum ionizing particles
(MIPs).  Details of the detector design, construction, performance,
along with other relevant information regarding our experimental
method, are presented at the experiment Web site~\cite{pb_web}, and in
Ref.~\cite{Frl03a}.

Two sets of fast analog triggers accepted nearly all non-prompt pion
decay events and a substantial fraction of muon decays, for energy
depositions in the calorimeter exceeding a high threshold of ${\rm
HT}\simeq 51.7\,$MeV, or a low threshold of ${\rm LT}\simeq 4.5\,$MeV.
A pion stopping in the target initiates a 180$\,$ns long pion gate
$\pi$G, timed so as to sample pile-up events up to $30\,$ns prior to
the $\pi^+$ stop pulse.  Event triggers were generated on the basis of
a coincidence of the $\pi$G gate, a beam anti-prompt veto, and a
shower signal in the calorimeter.

Charged particles emanating from the target were tracked by two
cylindrical multi-wire proportional chambers~\cite{Kar98} and a
20-piece thin plastic scintillator veto hodoscope (PV).  The PV
signals provided discrimination between the background protons and
MIPs.  MIP detection efficiencies were continuously monitored.  The
average efficiencies of the inner and outer MWPCs were $\epsilon_{\rm
C_1}=93.7$\% and $\epsilon_{\rm C_2}=97.9$\%, respectively.  The
average efficiency of the PV hodoscope was $\epsilon_{\rm PV}=98.9$\%.
The overall inefficiency for distinguishing neutral and charged
particles was therefore $1.5\cdot 10^{-5}$.

$\pi\to e\nu\gamma$ candidate events were selected from the two-arm HT
and prescaled one-arm HT data sets by requiring one neutral shower in
the calorimeter in coincidence with a positron track.  In cases where
there was more than one possible coupling (i.e., more than one neutral
or positron track), the pair most nearly coincident in time was
chosen.  Our data set covers three kinematic regions:
\begin{itemize}
\item[$A$:] $E^{\text{cal}}_{e^+},E^{\text{cal}}_\gamma > 
                51.7\,$MeV (two-arm HT), 
\item[$B$:] $E^{\text{cal}}_{e^+} > 20.0\,$MeV, 
              $E^{\text{cal}}_{\gamma} > 55.6\,$MeV (one-arm HT),
\item[$C$:] $E^{\text{cal}}_{e^+} > 55.6\,$MeV, 
              $E^{\text{cal}}_{\gamma} > 20.0\,$MeV (one-arm HT),
\end{itemize}
where the superscript ``cal'' refers to values measured in the CsI
calorimeter.  In all three regions the relative angle
$\theta^{\text{cal}}_{e^+\gamma} >40.0^\circ$.

\begin{figure} [b] 
\includegraphics[scale=0.75]{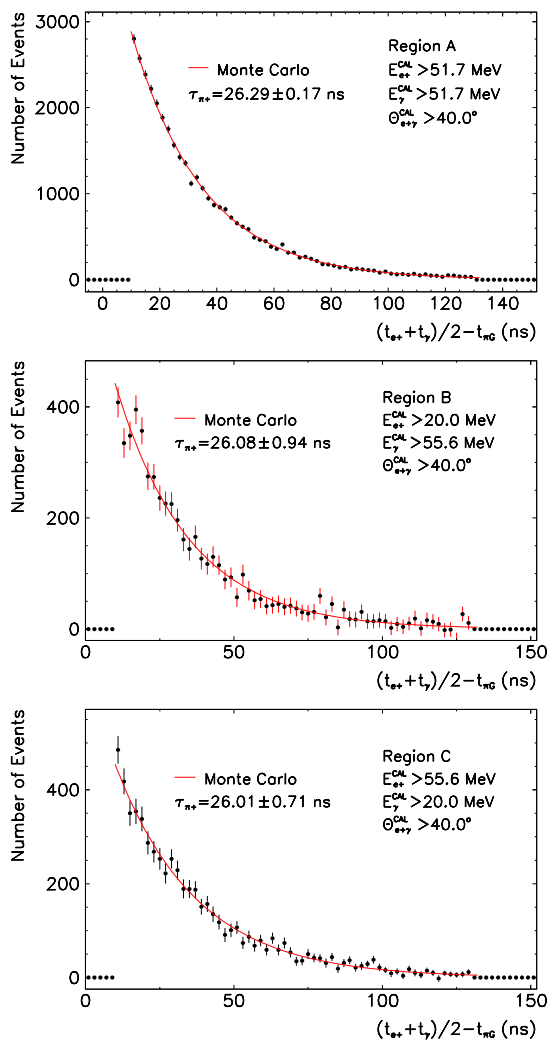}
\caption{$\pi^+\to e^+\nu\gamma$ event timing spectra after
subtraction of accidental background events.  Predicted Monte Carlo
decay functions are shown as full lines.}
\label{rp_tim}
\end{figure}

The signal-to-background ratios $S/B$ are shown in Fig.~\ref{rp_sn}
where the time differences between the radiative photon and the
positron tracks are histogrammed.  The high energy pairs of class $A$
show virtually no background ($S/B\ge 300$), while the class $B$ ($C$)
events with lower energy positrons (photons) have $S/B= 3.8$ (7.6).

We define the coincidence time window by $\Delta t_{\rm I}\equiv \vert
t_{e^+}-t_\gamma\vert \le$\,5\,ns. The accidental background was
sampled in two sidebands $\Delta t_{\rm O}$, $-10\,{\rm ns}<
t_{e^+}-t_\gamma <-5\,{\rm ns}$ and $5\,{\rm ns}< t_{e^+}-t_\gamma
<10\,{\rm ns}$.

The accidental background is dominated by positrons from $\pi\to e\nu$
or the $\pi$-$\mu$-$e$ decay chain, accompanied by an unrelated
accidental neutral shower.  This background was removed by subtracting
histograms of observables projected using the out-of-time cut $\Delta
t_{\rm O}$ from the in-time histograms projected via the $\Delta
t_{\rm I}$ cut.

In addition to shower energies, we also measured the directions of the
positron and photon, thus overdetermining the final three-body state.
The positron direction was fixed using MWPC hits, while the photon
direction was reconstructed from the pattern of hits in the
calorimeter.  Events with kinematics incompatible with the $\pi\to
e\nu\gamma$ decay were rejected in the final data sample.

The time distributions of the $\pi\to e\nu\gamma$ events with
respect to the $\pi^+$ stop time are plotted in Fig.~\ref{rp_tim}.
The fitted exponential curves are consistent with the $\pi^+$ lifetime
and demonstrate the purity of the final data set.

Non-accidental background sources are (1) $\pi\beta$ events for which
one $\pi^0$ decay photon converts in the target, producing a charged
track in the detector and (2) two-clump showers originating from a
single $\pi\to e\nu$ positron when a secondary shower photon or
positron interacts in the calorimeter far enough from the primary hit
to appear as a separate clump (``split-clump'' events).

Starting from the measured net yield of the $\pi\beta$ decay events,
we have used the Monte Carlo (MC) simulation of $\gamma$ conversions
in the inner detector to determine that class (1) background events
contribute 12.9\,\% of the signal in the kinematic region $A$ and
4.3\% (3.8\,\%) in region $B$ ($C$).  These $\pi\beta$ contaminations
were subtracted in the calculation of $\pi\to e\nu\gamma$ yields.  For
the clump energy threshold $E^{\text{cal}}_{e^+,\gamma}>20.0$\,MeV
used in the analysis the ``split-clump'' background (2) can be
neglected.

\begin{figure} [t] 
\includegraphics[scale=0.55]{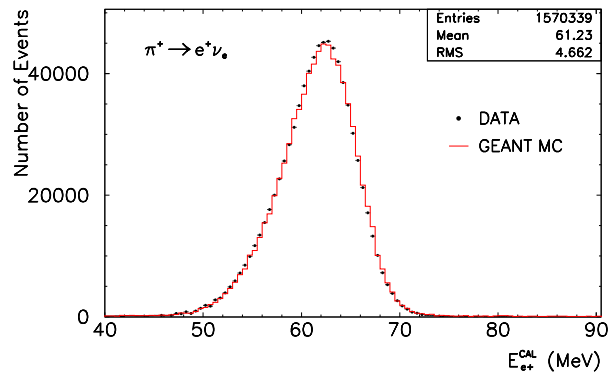}
\caption{Background-subtracted $\pi^+\to e^+\nu$ energy spectrum
taken with the one-arm trigger. The {\tt GEANT}-simulated
detector response is represented by the full line histogram.}
\label{p2e_en}
\end{figure}

We have normalized the radiative pion decay rate to the total
branching ratio of the $\pi\to e\nu$ ($\pi e2$) events which were
recorded in parallel using the prescaled one-arm calorimeter trigger.
The energy spectrum of the $\pi^+\to e^+\nu$ positrons is shown in
Fig.~\ref{p2e_en}.  This procedure assures that most factors in the
normalization, including the total number of decaying $\pi^+$'s and
the combined tracking efficiency of $e^+$'s largely cancel out.

The absolute branching ratio for the $\pi^+\to e^+\nu\gamma$ decay
can be calculated from the expression:
\begin{equation}
  {R}_{\pi e2\gamma}=
   \frac{N_{\pi e2\gamma}\, p_{\pi e2\gamma}}
        {N_{\pi^+}\, g_\pi\, A_{\pi e2\gamma}\, \tau_{l}\,
         \epsilon_{\rm PV}\,\epsilon_{\rm C_1}\,\epsilon_{\rm C_2}}\,,
                                                           \label{Eq:4}
\end{equation}
where $N_{\pi e2\gamma}$ is the number of the detected $\pi\to
e\nu\gamma$ events, $p_{\pi e2\gamma}$ is the corresponding prescaling
factor (if any), $N_{\pi^+}$ is the number of the decaying $\pi^+$s,
$g_\pi=\int_{t_1}^{t_2}\exp({-t/\tau_{\pi^+}})dt$ is the $\pi^+$ gate
fraction, $A_{\pi e2\gamma}$ is the detector acceptance incorporating
the appropriate cuts, and $\tau_l$ is the detector live time fraction.
An analogous expression can be written for the total $R_{\pi e2}$
branching ratio.  Combining the two, we obtain
\begin{equation}
   R_{\pi e2\gamma} = R_{\pi e2}
     \frac{N_{\pi e2\gamma}\,p_{\pi e2\gamma}}
            {N_{\pi e2}\,p_{\pi e2}}\,
     \frac{A_{\pi e2}}{A_{\pi e2\gamma}}\,
     \frac{\epsilon_{\pi e2}}{\epsilon_{\pi e2\gamma}}\,,
                                                         \label{Eq:5}
\end{equation}
where the $\epsilon$'s denote the properly weighted products of
$\epsilon_{\rm PV}\cdot\epsilon_{\rm C_1}\cdot\epsilon_{\rm C_2}$ for
the two data sets.  The detector acceptance ratio takes on values
between 0.057 and 0.209, depending on the kinematic region.

The experimental acceptances depend on both the detector response and
the decay amplitudes and are calculated in a Monte Carlo (MC)
simulation.  The {\tt GEANT}~\cite{Bru94} simulation of the PIBETA
detector response included: (i) the detailed geometry of the active
detectors and the passive support material, (ii) the measured detector
energy and timing responses, (iii) event generators for $\pi$ and
$\mu$ decays including measured accidental pile-up rates, and (iv) the
photo-absorption reactions in the CsI calorimeter incorporated in the
{\tt GEANT} code.

In the Standard Model the differential rate of the $\pi^+\to
e^+\nu\gamma$ decay can be written in the form~\cite{Bry82}:
\begin{align}
  & \frac{d\Gamma_{\pi e2\gamma}}{dx\,dy} = 
      \frac{\alpha}{2\pi} \Gamma_{\pi\text{e2}}
        \Big\{ IB\left( x,y \right) 
       + \left( \frac{F_V m_\pi^2}{2 f_\pi m_e} \right)^2 \nonumber \\
  & \times \big[ \left( 1+\gamma \right) ^2 SD^+ \left(x,y \right)
      + \left( 1-\gamma \right) ^2 SD^- \left(x,y \right) \big]
                                                    \label{Eq:6} \\
  &  +\left( \frac{F_V m_\pi}{f_\pi} \right) 
      \left[\left( 1+\gamma \right)S_{\rm int}^+\left( x,y \right) +
      \left( 1-\gamma \right) S_{\rm int}^- \left( x,y \right) \right]
       \Big\}\,,                                 \nonumber
\end{align}
where the $IB$, $SD^\pm$, and $S_{\rm int}^\pm$ (IB--SD interference)
terms depend on the kinematic variables $x=2E_\gamma/m_\pi$ and
$y=2E_e/m_\pi$;  $E_e$ and $E_\gamma$ are the physical (``thrown'')
energies.

We took the PDG value $\Gamma(\pi e2)/\Gamma_{\rm total}=
1.230(4)\cdot 10^{-4}$ \cite{PDG} and calculated the experimental
branching ratios using the {\tt MINUIT} least chi-square
program~\cite{Jam89}.  We have added integral radiative corrections of
$-1.0\,$\% in region $A$, $-1.4\,$\% in $B$, and $-3.3\,$\% in $C$, to
the theoretical $R_{\text{the}}$'s~\cite{Kur03}.  The minimization
program simultaneously fits two-dimensional $(E^{\text{cal}}_{e^+},
E^{\text{cal}}_{\gamma})$ distributions in all three phase space
regions, constraining the integrals to the experimental branching
ratio values ($R_{\text{exp}}$).

The acceptance ${A_{\pi e2\gamma}}$ is recalculated in every iteration
step of our analysis with the cuts applied to the physical
(``thrown'') energies, $E$, following cuts applied to measured
particle energies and angles, $E^{\text{cal}},\theta^{\text{cal}}$.
Hence, our experimental branching ratios can be compared directly with
theoretical absolute decay rates.

The statistical uncertainties of the experimental yields are 0.6\,\%,
1.7\,\% and 1.5\,\% for the regions $A$, $B$, and $C$, respectively.
Systematic uncertainties of 1.8\,\% for region $A$, dominated by
$\pi\beta$ background subtraction, 2.3\,\% and 3.1\,\% for regions $B$
and $C$, respectively, both dominated by acceptance ratio
uncertainties, were added in quadrature.

\begin{figure} [t] 
\includegraphics[scale=0.42]{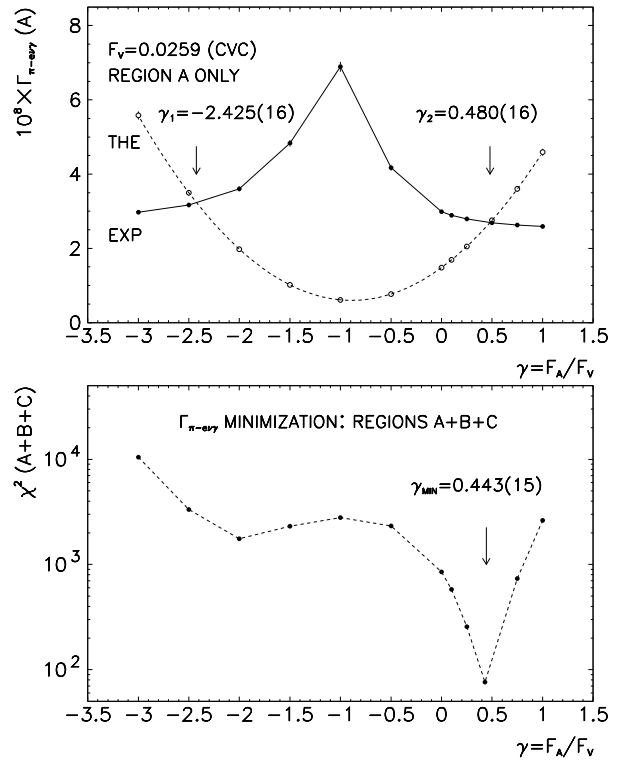}
\caption{Top: $\pi^+\to e^+\nu\gamma$ branching ratio values as a
function of $\gamma\equiv F_A/F_V$.  The theoretical parabola follows
from the $V - A$ model, Eq.~(\protect{\ref{Eq:6}}).  The experimental
values reflect fits to region $A$ data only.  Bottom: minimum $\chi^2$
values of simultaneous fits to the entire data set (regions $A$, $B$,
$C$).}
\label{fig:rp_br}
\end{figure}

The dependence of the region-$A$ experimental and theoretical
branching ratio on the value of $\gamma$ is shown in
Fig.~\ref{fig:rp_br} (top), indicating two solutions.  The positive
$\gamma$ solution is preferred by $\sim\,50$:1 once data from regions
$B$ and $C$ are included in the analysis (bottom plot).  We compare
the experimental and theoretical branching ratios for the three phase
space regions in Table~\ref{tab1}.  We note that due to the large
statistical and systematic uncertainties present in all older
experiments, our values are consistent with previously published
measurements.  The best CVC fit to our data yields $\gamma=0.443\pm
0.015$, or $F_A=0.0115(4)$ with $F_V \equiv 0.0259$.  This result
represents a fourfold improvement in precision over the previous world
average, and is consistent with chiral Lagrangian
calculations~\cite{Hol86,Bij97,Gen03}.

In summary, our experimental $\pi^+\to e^+\nu\gamma$ branching
ratios and energy distributions in kinematic regions $A$ and $C$ are
compatible with the ($V-A$) interaction.  The sizable 19\,\% shortfall
of the measured branching ratio compared to the theoretical one in
region $B$ dominates the total $\chi^2$, and is disconcerting.  Thus,
in a fit restricted to region $A$ data only, we obtain $\gamma =
0.480\pm 0.016$, the same as when analyzing the combined data from
regions $A$ and $C$.  Significantly, all previous studies save one
(which, too, found an anomaly~\cite{Bol90}), have analyzed only data
with kinematics compatible with our region $A$.

Taking the $(V-A)$ interaction and the CVC hypothesis as valid, the
deficit could be caused by a specific detection or analysis
inefficiency in our experiment, appearing dominantly in region $B$.
Detailed cross-checks, including absolute total $\pi\to e\nu$,
$\pi^+\to\pi^0e^+\nu$, $\mu\to e\nu\bar{\nu}$, as well as partial
$\mu\to e\nu\bar{\nu}\gamma$ branching ratio evaluations from our data
have thus far categorically excluded such inefficiencies
\cite{Frl03b}.  Alternatively, the deficit could be caused by an
inadequacy of the present $V-A$ description of the radiative pion
decay, i.e., Eqs.~(\ref{Eq:2}), (\ref{Eq:3}) and (\ref{Eq:6}) along
with the radiative corrections, or by an anomalous, non-$(V-A)$
interaction~\cite{Bol90,Her94,Che97,Chi93}.  We will focus on the
region $B$ discrepancy and the latter possibility in a separate
forthcoming Letter, noting that our results clearly call for further
theoretical and experimental work.

We thank W.~A.~Stephens, Z.~Hochman, and the PSI experimental support
group for invaluable help in preparing and running the experiment.  We
thank M.~V.~Chizhov and A.~A.~Poblaguev for fruitful discussions and
comments, as well as E.~A.~Kuraev and Yu.~M.~Bystritsky for
communicating to us their radiative correction results prior to
publication.  The PIBETA experiment has been supported by the
U.S. National Science Foundation, the U.S. Department of Energy, the
Paul Scherrer Institute, and the Russian Foundation for Basic
Research.

\begin{table} [t]
\caption{Best-fit $\pi\to e\nu\gamma$ branching ratios obtained with
$F_V=0.0259$ (fixed) and $F_V = 0.0115(4)$ (fit);
$\chi^2/\text{d.o.f.} = 25.4$.  Measured ($R_{\rm exp}$) and
theoretical ($R_{\rm the}$) branching ratios are shown for the three
indicated phase space regions.  Radiative corrections are included in
the calculations.}
\label{tab1}
\begin{ruledtabular}
\begin{tabular}{ccccc}
 $E^{\text{min}}_{e^+}$ & $E^{\text{min}}_\gamma$ 
                 & $\theta^{\text{min}}_{e\gamma}$  
                                & $R_{\rm exp}$ & $R_{\rm the}$ \\
 (MeV) & (MeV) &       & $(\times 10^{-8})$ & $(\times 10^{-8})$  \\
\hline
 $50$  & $50$  & $-$        & $2.71(5)$ & $2.583(1)$ \\
 $10$  & $50$  & $40^\circ$ & $11.6(3)$ & $14.34(1)$ \\
 $50$  & $10$  & $40^\circ$ & $39.1(13)$ & $37.83(1)$  \\
\end{tabular}
\end{ruledtabular}
\end{table}

\bibliography{pienug20}

\end{document}